\documentstyle[12pt,epsfig]{article}

\begin{document}

  \newcommand{\ccaption}[2]{
    \begin{center}
    \parbox{0.85\textwidth}{
      \caption[#1]{\small{{#2}}}
      }
    \end{center}
    }

\hfill {\sf IHEP--97--66}

\vspace{0.5cm}

{\Large { \sf 

{\bf On Mechanism of Charmed $c$--Quarks Fragmentation in Hadronic Collisions 
}}} 

\vspace{1cm}

\begin{center} 
A.K.~Likhoded$^1$ and S.R.~Slabospitsky$^2$ 
\end{center}

\vspace{1cm}

\begin{center}

State Research Center \\
Institute for High Energy Physics, \\
Protvino, Moscow Region 142284 \\
Russia

\end{center}

\vspace*{1.5cm}

We propose a modification of the mechanism of charmed quarks fragmentation 
into $D$ mesons in hadronic collisions. It is shown that the distinction in
valence quark distributions in the initial $\pi^{\pm}$ and $K^{\pm}$ mesons
leads to different inclusive spectra of $D$ and $D_s$ mesons produced in 
$\pi^{\pm}$ and $K^{\pm}$ beams.

\vspace*{1cm}

\rule{3cm}{0.5pt}

$^1$E--mail:~LIKHODED$@$mx.ihep.su,

$^2$E--mail:~SLABOSPITSKY$@$mx.ihep.su

\newpage

\section{\bf INTRODUCTION } 

Hadroproduction of the particles with open charm, among other reactions 
($e^+ e^- \to c \bar c$ etc) is of interesting possibility to study a
hadronization process of the heavy quarks, such as 
$c \to D, \Lambda_c, ...$~\cite{1}.

It is well known that the production of charmed particles in
$e^+ e^-$--anni\-hi\-la\-tion, 
\begin{eqnarray}
 e^+ \, e^- \, \, \to \, \, D \, X, \label{1} 
\end{eqnarray}
may be separated into a process of $c \bar c$--pair production and a process
of independent fragmentation of each $c$ ($\bar c$) quark into $D$--meson:
\begin{eqnarray}
 c (\bar c) \, \, \to \, \, D (\bar D) \, X. \label{2} 
\end{eqnarray}

In so doing the fragmentation process is described by means of fragmentation
function $D(z)$, where $z = p_D/p_c$ ($p_D$ and $p_c$ are the momenta of 
$D$--meson and $c$--quark, respectively). 

It worth to note that a scale--invariant description becomes available at
sufficiently high energies, i.e. in the limit of $\sqrt{s} \gg m_D$. For the
low energy case ( $\sqrt{s} \ge m_D$) the role of non--scaling (power)
corrections is very essential. As a result at that energy region one can not 
represent the cross section in a simple factorized form:
\begin{eqnarray}
 \frac{d \sigma_D}{dz} \approx \sigma_{c \bar c} \otimes D(z). \label{3} 
\end{eqnarray}

The question arises of what extent a mechanism of fragmentation of heavy
quarks, produced in $e^+ e^-$ annihilation, can be applied for the case of
hadronic interactions ?

Note, that in hadronic interactions the presence of light partons from the
initial hadrons will affect the hadronization scenario. One should distinguish
two kinematical regions, namely, the region of 
$p_{\top} < p_0$ and the region of $p_{\top} > p_0$. 
Like in $e^+ e^-$ annihilation, the fragmentation process above $p_0$ 
has a fragmentation character and the cross section has the form as follows:
\begin{eqnarray}
 \frac{d \sigma_D}{d p_{\top}} = \int 
 \frac{d \sigma_c}{d k_{\top}} \otimes D(z) dz. \label{4}
\end{eqnarray}
At the region of $p_{\top} < p_0$ the presence of light partons, produced 
simultaneously with $c \bar c$ pair, may radically altered the factorization
in the form of~(\ref{4}). 

In our previous publication \cite{ls97} we made an effort to take into account
this fact by consideration of a some contribution into hadronization as a 
process of recombination 
with valence quarks. The rest part is described by a fragmentation mechanism.
In the framework of such a consideration one can properly describe the
asymmetry in the production of leading and non-leading charmed 
hadrons~\cite{ls97}. Note, that our description of total spectrum is not well
adequate. It explains because the excess of our calculated distribution 
at low $x$. This fact
is caused by the application of the fragmentation model for the whole
kinematical region. 

In the present article we make an effort to improve the description of 
$c$--quark hadronization at low $p_{\top}$ region. We try do not use the
fragmentation model as a basic mechanism. We apply this model only at
boundaries of the phase space, where $c$--quark momenta achieve their maximum
values and where the hadronic accompaniment of $c$--quarks is very small. 

We note also, that in the present article we do not consider the question
related to an absolute magnitude of the cross section of charmed  particle
production. The matter is that such a problem is determined by taking into
account the higher order corrections of the perturbative QCD
as well as the choice of the strong coupling constant of $\alpha_s(\mu^2)$.
Indeed, the analysis of $O(\alpha_s^3)$ corrections to $c \bar c$ pair
production cross section (see, for example,~\cite{1}) exhibits that their
inclusion practically do not alter the form of the inclusive distributions of
$c$--quarks. Therefore, in what follows we shall restrict our consideration to
study the form of differential spectra of the charmed particles.

The article is organized as follows. We consider the modification of the
fragmentation mechanism in the Section~2. We compare in Section~3 the results
of our calculations with the experimental data on $D$--meson production in 
$\pi$--beam. Section~4 presents the model predictions for the $K$ beam. The
main results are summarized in Conclusion. 

\section {\bf $c$--QUARK HADRONIZATION IN HAD\-RO\-NIC COLLISIONS }

In the parton model, the cross section for the production of heavy quarks $Q$
in hadron--hadron collisions has the form
\begin{eqnarray}
\sigma (h_1 h_2 \to Q \bar Q X) =  \Sigma_{i,j} \int \hat 
\sigma(i j \to Q \bar Q)  f^{h_1}_{i}(x_1) d x_1 f^{h_2}_j(x_2) dx_2, 
\label{f1} 
\end{eqnarray}
where summation is performed over all partons participating in $Q$ quark
production $i j \to Q \bar Q$, $\hat \sigma$ is the cross section for the
corresponding hard subprocess, $f^h_i(x)$ is the distribution of $i$--type
partons in the hadron $h$, $x_{1(2)}$ is the fraction of the initial-hadron
momentum carried away by the corresponding parton.

As it mentioned in the Introduction, the presence of light partons from the
initial hadrons leads to radically different scenarios of the hadronization of
charmed $c$--quarks, produced in hadron--hadron collisions and in 
$e^+ e^-$ annihilation. In particular, the interaction in final state with 
valence quarks from initial hadrons (recombination) makes possible to
explain the leading particle effect in the charmed hadrons 
production~\cite{ls97,old} (i.e. the difference in $x$--distributions of
$D$ and $\bar D$ mesons, as well as $\Lambda_c$ and $\bar \Lambda_c$
baryons).

Such an inclusion of interaction of charmed quarks with valence quarks from
the initial hadrons is carried out by the introduction of the quark
recombination function of $R(x_V, z; x)$~\cite{old}. Below we present the
basic concepts of the recombination mechanism. The detail description of this
mechanism is given elsewhere~\cite{ls97,old}. The recombination of $q_V$ and 
$\bar Q$ quarks into $M_Q$ meson is described by the function of 
$R(x_V, z; x)$:
\begin{eqnarray}
 R(x_V, z; x) = \rho(\xi_V, \xi_Q) \;\; \delta(1 - \xi_V - \xi_Q), 
\label{rec1} \\ 
  \rho(\xi_V, \xi_Q) \; = \; \frac{\Gamma(2 - \alpha_V-\alpha_Q)}
 {\Gamma(1-\alpha_V) \Gamma(1-\alpha_Q)}
 \xi_V^{(1-\alpha_V)} \xi_Q^{(1-\alpha_Q)}, \label{rec2} 
\end{eqnarray}
where $\xi_V = x_V / x$ and $\xi_Q = z / x$, while $x_V$, $z$, and $x$ are the
fractions of the initial-hadron c.m. momentum that are carried away by the
valence quark, quark $\bar Q$, and the meson $M_{\bar Q}$, respectively.
$\alpha_V$ and $\alpha_Q$ are the intercepts of the leading Regge-trajectories
for the $q_V$ and $\bar Q$ quarks, respectively. In our calculations we 
use~\cite{collin,klp}:
\begin{eqnarray}
\alpha_u = \alpha_d = \frac{1}{2}, \quad \alpha_s \approx 0, \quad
  \alpha_c \approx -2.2. \label{rec22}
\end{eqnarray}

With the aid of the function $R(x_V, z; x)$ describing the recombination of
the quarks $q_V$ and $\bar Q$ into a meson, we represent the corresponding
contribution to the inclusive spectrum of $M_{\bar Q}$ mesons as follows:
\begin{eqnarray}
x^{\ast} \frac{d \sigma^{rec} }{dx} = R_0 \int x_V z^{\ast} 
\frac{d^2 \sigma}{dx_V dz} \; 
R(x_V, z ; x) \frac{dx_V}{x_V} \frac{dz}{z}, 
\label{sig1} 
\end{eqnarray}
where $x^{\ast} = 2 E / \sqrt{s}$ and $x = 2 p_l / \sqrt{s}$ (here, $E$ and
$p_l$ are the energy and longitudinal momentum of the $M_{\bar Q}$ meson in
the c.m.s. of the initial hadrons); $x_V$  and $z$ are the momentum fractions
carried away by the valence quark and heavy antiquark, respectively, 
and $x_V z^{\ast} \frac{d^2 \sigma}{d x_V \; dz}$ is the double-differential
cross section for the simultaneous production of the quarks $q_V$ and
$\bar Q$ in a hadronic collision. 

The parameter, $R_0$, is the constant term of the model, that determines the
relative contribution of recombination. In the present article the best
description of the experimental data could be achie\-ved at
\begin{eqnarray}
R_0 \approx 0.8.  \label{sigr}
\end{eqnarray}

Note, that  using of the recombination with the valence quarks is 
indispensable to an explanation of the leading particle effect. At the same 
time
its contribution in the total inclusive cross section production of the
charmed particles is sufficiently small~($\sim 10\%$).  This mechanism
dominates in the high $x$ region.

In the conventional fragmentation mechanism the inclusive cross section for
the charmed hadrons ($D$ mesons) production has the form as follows:
\begin{eqnarray}
E_H \frac{d^3 \sigma^F}{d^3 p_H} \; = \; \int 
E_c \frac{d^3 \sigma(h_1 h_2 \to c X)}{d^3 p_c} \; D(z)
\, \delta(\vec p_H \, - \, z \vec p_c) \, d^3 p_c. \label{f2}
\end{eqnarray}

In the low $x$ region, which determines the basic contribution into the total
cross section for charmed hadron production, the hadronization process of $c$
quark has more complicated nature.

Indeed, when evaluating the spectra of charmed particles produced in hadronic
collisions one assumes that the fragmentation function 
$D(z)$ is well known from other experiments (in particular, from
$e^+ e^-$ annihilation).
One of the widely used parameterization is as follows \cite{peters}:
\begin{eqnarray}
  D(z) \sim [z( 1 - \frac{1}{z} - \frac{\varepsilon}{1-z} ) ] ^{-2}, \label{f4}
\end{eqnarray}
where the parameter $\varepsilon \approx m^2_q / m^2_Q$ is determined by the 
type
of a hadron (for instance, one has $\varepsilon_{D^0} = 0.135 \pm 0.010$ and
$\varepsilon_{D^{\ast}} = 0.078 \pm 0.008$~\cite{pdg}).

Another parameterization, proposed by us early~\cite{klp}, takes into account
the Regge-asymptotic at $z \to 0$: 
\begin{eqnarray}
  D(z) \sim z^{-\alpha_Q} ( 1 - z)^{\gamma}, \label{f5}
\end{eqnarray}
where $\gamma \approx 1$, while $\alpha_Q$ is the intercept of the leading
Regge-trajectory for the $Q$ quark ($\alpha_c \approx -2.2$,
see~(\ref{rec22})).

Both of these parameterizations provide a reasonably fair description of the 
experimental
data. The comparison of description of the reaction 
$e^+ e^- \to D X$ by means of the fragmentation functions in both form of 
(\ref{f4}) and (\ref{f5}) is presented in~\cite{expfr1,expfr2}. 

As it mentioned in Introduction, the use of the fragmentation function is
justified at asymptotically large values of the invariant mass of 
$c \bar c$ pair, namely at $M_{c \bar c} \; \gg 2 m_c$. However, this
condition is not obeyed in the hadronic production of the charmed particles.
For such a case the principal contribution into inclusive charm production
cross section results from $c$ quarks with low values of the invariant mass of
$c \bar c$ pair ($M_{c \bar c} \; \ge 2 m_c$). These quarks dominate in the
central region on Feynman variable of $x$. 
The pairs of $c$ quarks with a large invariant mass, where one may apply the
fragmentation formalism, give a noticeable  contribution at high $x$ and high
$p_{\top}$, as well.

These arguments are illustrated in Fig.~1. This picture presents the inclusive
$x$--distribution for $c$ quark for all $M_{c \bar c}$ (the upper histogram) 
and
for the $c$ quarks with the invariant mass of $M_{c \bar c} \geq M_0 = 10$~GeV
(the lower histogram). As is seen from the figure just the charmed quarks with
small invariant masses of $c \bar c$ pair give the dominant contribution in
the charm production cross section in the central region, while the region of
$x \to 1$ corresponds to the contribution due to a large  values of mass of
$M_{c \bar c}$.

In this figure we also present the spectrum of charmed particles summed over
all types $D$ and $\bar D$ mesons (for the reaction
of $\pi^- N$ collisions at $E_{\pi} = 250$~GeV~\cite{exp1}.) 
This experimental spectrum should be compared with the distribution of $c$
quarks.
One may deduces from the Fig.~1 that the correspondence, like a duality
relation, takes place. Namely, the spectrum of charmed hadrons, summed over
all types of charmed mesons, is well described by the inclusive spectrum of
$c$-quarks.

Such a satisfactory description of summed spectra of $D$ mesons by pure 
$c$-quark spectra was also pointed out 
early~\cite{1,exp1}. However, it is evident
that in the framework of pure fragmentational mechanism one should expect the
identical spectra of both $D$ and $\bar D$ mesons (as well as charmed baryons
and antibaryons). As a result, one can not reproduce the leading particle
effect. As it mentioned above, this effect can be described with the help of
the recombination mechanism. Therefore, one could assume that the inclusive $D$
meson production cross section is described by the sum of two mechanisms as
follows: 
\begin{eqnarray}
 \frac{d \sigma_D}{dx} =  \frac{d \sigma^{HF}_D( \vec p_D = \vec p_c)}{dx} 
 + \frac{d \sigma^{rec}_D}{dx}, \label{sig2}
\end{eqnarray}
where the first term corresponds to the "hard" (HF) fragmentation  (in that
mechanism a charmed quark does not lose its momentum in hadronization
process), while the second term corresponds to recombination contribution
(that mechanism takes into account the charmed $c$-quark interaction with
valence quarks from initial hadrons).

However, such a simple addition of the recombination contribution to the $c$
quark spectrum (i.e. $D$ meson) does not provide to reproduce the behavior of
$x$--dependence of the corresponding asymmetry~$A$:
\begin{eqnarray}
A = \frac{ 
 \frac{d \sigma}{dx}(leading) \; - \; \frac{d \sigma}{dx}(non-leading) }
 {\frac{d \sigma}{dx}(leading) \; + \; \frac{d \sigma}{dx}(non-leading) }. 
\label{sig8}
\end{eqnarray}

Indeed, the Fig.~2 presents the description of the asymmetry (\ref{sig8}) by
means of the equation~(\ref{sig2}). The different histograms in this figure
correspond to the different values of the parameter $R_0$ (see the
equation (\ref{sig1}) for the recombination mechanism). As
is seen from this figure it is impossible to obtain the simultaneously proper
description of the asymmetry $A$ both at small $x$ ($0 \le x \le 0.4$) and at 
high $x$  ($0.5 \le x \le 0.8$) as well as with the description of
the $D$ mesons inclusive $x$--distributions.

This is because the simple equating of $D$ meson spectrum to the $c$ quark
spectrum (i.e. $D(z) \sim \delta(1-z)$) leads to exclusively "hard" spectra of
$D$ mesons at high $x$. Therefore, we get the conclusion that one needs to use
a more softer (in compare with $\delta (1-z)$) fragmentation function for the
description of charmed quark hadronization at high values of the Feynman
variable of $x$.

Bearing in mind the preceding, we consider the modification of the
conventional fragmentation scenario of charmed quarks hadronization. It seem
likely that in the region of small invariant mass of $c \bar c$ pair the
description of the hadronization in terms of fragmentation function
(evaluating from the $e^+ e^-$ annihilation) is not justified. Indeed, one has
large amount of partons from initial hadrons in the central region of $x$.
Therefore, the $c$ quark in combination with one of such a parton can easily
to produce a charmed hadron. Such a process occurs practically without any
loss of $c$ quark momentum (i.e. $\vec p_D \approx \vec p_c$). Therefore, in
the small $x$ region one should expect the coincidence of the spectra of $D$
mesons and $c$ quarks. Whereas at high $x$ region one may use the conventional
 fragmentation mechanism (and a recombination mechanism, as well).

Hence, we consider two regimes of the charmed quark fragmentation.

 a) Close to the threshold of $c$ quarks production, i.e. 
at $M_{c \bar c} \geq 2 m_c$, the momentum of the produced $D$ meson should
practically coincides with the momentum of the charmed parent-quark.

 b) For $c \bar c$ pair with the invariant mass $M_{c \bar c}$ greater that 
certain scale of $M_0$ (where $M_0 \gg 2 m_c$) the $c$ quark hadronization
process may be described with the help of fragmentation function (for
instance, in the form of~(\ref{f4}) or~(\ref{f5})).

In terms of the fragmentation mechanism these two regimes may be represented
in a uniform way by introduction of the dependence of fragmentation function 
from an invariant mass of $c \bar c$ pair as follows:
\begin{eqnarray}
 D^{MF}(z, M_{c \bar c}) = \left \{ \begin{array}{l c} 
 \sim \delta(1-z) & {\rm at } \; \; M_{c \bar c} \approx 2m_c \\
 D(z) \; \; {\rm from } \; (\ref{f4}) \; {\rm or } \; (\ref{f5}) & 
 {\rm at } \; \; M_{c \bar c} \geq M_0 \end{array} \right. \label{a1} 
\end{eqnarray}

It worth to note, that assumed $M_{c \bar c}$--dependence of the $c$-quark
fragmentation function has nothing to do with logarithmic violation of scaling
in the fragmentation function.

To reproduce in a uniform way the both two fragmentation regimes~(\ref{a1}) we
use the simplest expression for $D(z, M_{c \bar c})$ in the from of~(\ref{f5})
as follows:
\begin{eqnarray}
 D^{MF}(z, M_{c \bar c}) \sim z^{-\alpha(M_{c \bar c})} (1-z), \label{a2}
\end{eqnarray}
with two additional conditions on $\alpha(M_{c \bar c})$:
\begin{eqnarray}
 \begin{array}{l c c c l} 
 \alpha(M_{c \bar c}) & \to & -\infty & \qquad {\rm at} &
\quad M_{c  \bar c} \, \to \, 2m_c, \label{a3} \\ 
 \alpha(M_{c \bar c}) & \to & \alpha_c &\qquad {\rm at} &
\quad M_{c  \bar c} \, \approx \, M_0. \label{dt3}
 \end{array} 
\end{eqnarray}
Our parameterization for $\alpha(M_{c \bar c})$ is presented in the
Appendix~1. A fit to experimental data exhibits that the magnitude of the
parameter $M_0$,
\begin{eqnarray*}
 M_0 \; \approx \; 10 \; {\rm GeV},
\end{eqnarray*}
provides a satisfactory description of experimental results. Such a magnitude
does not contradict to the experiments in $e^+ e^-$ annihilation, where
already at the energy of $\sqrt{s} \approx 10$~GeV one can describe the
process in terms of fragmentation mechanism by using the dependence in the
form of~\ref{3}. 

Thus, the summed differential cross section production of 
charmed hadron ($D$-meson) can be represented as follows:
\begin{eqnarray}
 \frac{d \sigma_D}{dx} =  \frac{d \sigma^{MF}_D}{dx} + \; 
   \frac{d \sigma^{rec}_D}{dx}, \label{sig4}
\end{eqnarray}
where $\frac{d \sigma^{MF}_D}{dx}$ is the differential cross section 
production for $D$--meson due to $c$-quark
fragmentation (that is described by the equation (\ref{f2}) with the modified
fragmentation function of $D^{MF}(z, M_{c \bar c}))$, while 
$\frac{d \sigma^{rec}_D}{dx}$ is the differential cross section
production for $D$--meson resulted from $c$-quark recombination with valence 
quarks (see (\ref{sig1})). 

Like in the previous article\cite{ls97}, we assume that a mesonic state 
$(c \bar q)$ goes over into a vector $M_V$ or a pseudoscalar $M_{PS}$ meson
with the probability proportional to the spin factor:
\begin{eqnarray}
 M_{PS} \; : \; M_V \; = \; 1 \; : \; 3. \label{sig6}
\end{eqnarray}

\section {\bf COMPARISON OF THE MODEL PREDIC\-TIONS WITH THE EXPERIMEN\-TAL 
 RESULTS IN $\pi^-$ BEAMS } 

As it mentioned in the Introduction our article deals with the description of
two types of the inclusive $x$--distributions, namely, the differential cross
section for $D$ meson production (i.e. $\frac{d \sigma}{d x}$) and the
asymmetry $A(x)$.

Fig.~3 presents the description of the differential distribution 
$\frac{d \sigma}{d x}$ for the reaction
\[ 
 \pi^- \; N \; \to \; (D \; + \; \bar D) \; X.
\]
The experimental data is summed with respect to all types of $D$ meson. The
beam energy equals $E_{\pi} = 250$~GeV. 

As is seen from the figure our model (the modified fragmentation
$+$ recombination) reproduces satisfactorily the experimental data. Note,
also, that although the recombination contribution into the total cross
section is rather small  ($\le 10\%$), it plays a substantial role at high 
$x$ (see Fig.~3). In addition, one may deduce from this figure that pure
modified fragmentation can not provide the proper description of the inclusive
spectrum in the whole kinematic region.

Fig. 4 presents the description of the corresponding asymmetry (the leading
particle effect). In this case, too, the considered model provides the
description of the experimental data~\cite{exp1,exp2}. 

It should be particularly emphasized once more that the "hard" fragmentation 
(i.e. ֵ. $\vec p_D = \vec p_c$) makes it possible the description of the
differential spectrum in the whole kinematic region but can not reproduce the
$x$--dependence of the asymmetry.

In the Table~1 we compare our model predictions with the experimental data on
the total yields of the charmed mesons as well as with the predictions of the
Lund model~\cite{lund} (the experimental results and the Lund-model
predictions are taken from~\cite{exp3}). 

One may deduce from the Table~1 that our evaluations for the ratios of the
cross sections production of $D$ mesons are in agreement with the experimental
results within the experimental errors. Remind, that we obtain these results by
consideration of the two types of the hadronization of charmed $c$ quarks. The
behavior of these processes is determined by two parameters 
$R_0$ and $M_0$, as well as the distribution functions of the partons in the
initial hadrons.

The decisive test of the considered model should be the comparison of the
theoretical predictions with the experimental results in
$K^{\pm}$ and $\Sigma^-$ beams. The valence quarks distributions in these
hadrons differ substantially from those in $\pi^{\pm}$ and $p$ beams (see
below). Consequently, one should expect the different contributions due to
recombination mechanism in the inclusive spectra of charmed hadrons.

\section {\bf CHARM PRODUCTION IN THE BEAMS OF CHARGED $K^{\pm}$ MESONS } 

From the view of parton model the difference of $K^{\pm}$ mesons from 
$\pi^{\pm}$ mesons not only resides in the replacement of the valence 
$d$--quark by the strange valence $s$--quark. The distribution function of
valence quarks in $K^{\pm}$ mesons should change substantially, as well.

The simplest (without scaling violation) parameterization of the distribution
functions of the valence $q_1$ quark in the meson of $M(q_1 \bar q_2)$ with
the valence quarks of $q_1$ and $\bar q_2$ has the form as follows~\cite{old}:
\begin{eqnarray}
 V^{M(q_1 \bar q_2)}_{q_1} (x) = 
\frac{\Gamma(2 + \gamma_0 - \alpha_1 - \alpha_2)}
{\Gamma(1 - \alpha_1) \Gamma(1 + \gamma_0 - \alpha_2)}
 x^{-\alpha_1} (1 - x)^{\gamma_0 - \alpha_2}, \label{ka3}
\end{eqnarray}
where $\alpha_i$ is the intercept of the leading Regge-trajectory for the 
$q_i$ quark, while $\gamma_0$ is certain parameter.

The coefficient in the above equation is determined by the normalization
condition:
\[ 
 \int^1_0 V^M_q(x) d x = 1.
\]

The choice of the parameter of $\gamma_0$ in (\ref{ka3}) is governed by
asymptotic behavior of the structure functions at 
$x \to 1$. The quark counting rules predicts the value as follows:
\[
\gamma_0 - \alpha_2 = 1.
\]

From the well-known  asymptotic for $\pi$--meson,
\[ V^{\pi}(x)|_{x \to 1} \sim \frac{1}{\sqrt{x}} (1-x)^1 
\]
if follows that $\gamma_0 = \frac{3}{2}$. Taking into account that 
$\alpha_u = \alpha_d = 1/2$, while $ \alpha_s \approx 0$ (see (\ref{rec22})), 
we obtain the following distributions of the valence 
$u$ and $s$ quarks in $K^{\pm}$ meson: 
\begin{eqnarray}
 V^K_u \; & \sim & \; \frac{1}{\sqrt{x}} (1-x)^{3/2}, \label{ka4} \\
 V^K_s \; & \sim & \;  (1-x)^1. \label{ka5}
\end{eqnarray}
From the form of the distributions of (\ref{ka4}) and (\ref{ka5}) it is
evident that the valence $s$--quark in the $K$ meson is much "harder" than 
$u$ quark:
\begin{eqnarray*}
 < x^K_{s_v}> &=& 0.33, \\
 < x^K_{u_v}> &=& 0.166.
\end{eqnarray*}
From the given parameterization of the $K$ meson structure
functions it follows  that the total momentum fraction carried away by the
valence quarks is equal to:
\begin{eqnarray}
 < x^K_v> =  < x^K_{s_v}> + < x^K_{u_v}> = 0.5. 
\end{eqnarray}
Such a magnitude should be compared to the similar value for the $\pi$ meson:
\begin{eqnarray}
 < x^{\pi}_v> = < x^{\pi}_{d_v}> + < x^{\pi}_{u_v}> = 0.4.
\end{eqnarray}
We assume further that the distribution of the gluons in
$\pi^{\pm}$ and $K^{\pm}$ mesons are identical in form. The previous 
analysis~\cite{batun} of the evolution of structure functions of 
$\pi^{\pm}$ and $K^{\pm}$ mesons, where the evolution starts from the
different distributions of valence quarks, provides an argument in favor of
such an assumption. Other arguments in support of such an assumption are the
identical form of the spectra of charmed mesons produced in 
$\pi^{\pm}$ and $K^{\pm}$ beams~\cite{exp1}.

As mentioned above, the form of the distributions of the rest sea partons
in $K^{\pm}$--meson coincides with the form of corresponding parton
distributions in $\pi^{\pm}$ mesons: 
\[
f^K_{sea} (x) = \epsilon f^{\pi}_{sea} (x). 
\]
Here, the parameter $\epsilon$ takes into account the variation of the momentum
fraction carried away by the valence quarks in $K$ meson with respect to to
analogous value in $\pi^{\pm}$ meson:
\[
 \epsilon = \frac{ 1 -  < x^K_v>} { 1 -  < x^{\pi}_v>} \approx 0.8.
\]

Due to the different distributions of (\ref{ka4}) and (\ref{ka5}) we
should expect the different $x$--distributions of $D(c \bar u)$ and 
$D_s(c \bar s)$ mesons, produced 
in the $K^{\pm}$ beams (that resulted from the recombination with
valence quarks). Two--particle distributions of the
partons in $K$ meson, needed for such a calculation, can be easily evaluated
with the help of equation~(\ref{ka3}). Their explicit form is presented in the
Appendix~2.

Fig.~5 presents our model predictions for the leading particle effect in
$K^-$--meson beam at the energy of $E^K = 250$~GeV. As expected, our model
predicts the different $x$--dependence of the asymmetry for $D$ and $D_s$
mesons. For the strange--charmed $D_s$--mesons the leading particle effect is
more pronounced as compared with usual charmed mesons. The observation of
such a difference would be evidence in favor of the considered model.

Unfortunately, only the integral asymmetry magnitude is measured in the
experiment~\cite{expk}:
\begin{eqnarray}
 A^{exp}_K(D_s) = 0.25 \pm 0.11,
\end{eqnarray}
that should be compared with our prediction:
\begin{eqnarray}
 A^{theor}_K(D_s) = 0.29. 
\end{eqnarray}
As is seen our theoretical estimate is in agreement with the experimental
value.

\section{\bf CONCLUSION }

This article presents the "improved" model of charmed quark hadronization. Our
model provides the consistent description of the inclusive differential
spectra of $D$ mesons, produced in $\pi^- p $ collisions. Further step forward
in the understanding of a hadronization mechanism of heavy quarks would be
related with a consideration of the processes of charmed particle production
in $K$ and $\Sigma$ interactions. It caused, in particular, the different
distributions of valence quarks in $K$ and $\Sigma$ hadrons as compared with 
$\pi$ and $p$ beams. The next step of such an investigation should be a 
consideration of the process of charmed baryons production, where the diquarks
from the initial hadrons play a substantial role.

\vspace{0.8cm}

\noindent {\bf ACKNOWLEDGMENTS }

\noindent The authors thank V.G.~Kartvelishvili, V.V.~Kiselev, and M.~Mangano
for the fruitful discussions.

This work was supported in part by Russian Foundation for Basic Research,
projects no. 96--02--18216 and 96--15--96575.

\newpage

\newpage

\noindent {\bf APPENDIX~1. }

In order to deduce the parameterization of $\alpha(M_{c \bar c})$, which takes
into account the conditions~(\ref{a3}), i.e. $\alpha(2m_c) = \infty$ and
$\alpha(M_0) = \alpha_c$, we consider the expression for the first moment of
$\mu$ from the fragmentation function of $D^{MF}(z, M_{c \bar c})$ 
from~(\ref{a2}): 
\begin{eqnarray*}
 \mu(M_{c \bar c}) \, \equiv \, \int^{1}_{0} z D(z, M_{c  \bar c}) dz = 
\frac{ 1 - \alpha(M_{c \bar c})}{3 - \alpha(M_{c \bar c})}.
\end{eqnarray*}
The desired expression for $\alpha(M_{c \bar c})$ equals: 
\begin{eqnarray*}
 \alpha(M_{c \bar c}) = 
 \frac{ 1 - 3 \mu(M_{c \bar c})}{1 - \mu(M_{c \bar c})}.
\end{eqnarray*}
We assume the following (QCD--motivated) dependence for $\mu(M_{c \bar c})$ 
\begin{eqnarray*}
 \mu(M_{c \bar c}) \, = \, \left ( \frac{ \ln ( \frac {M_{c \bar c}}{2m_c}
 q_0)} {\ln q_0} \right )^d, 
\end{eqnarray*}
where $d \approx 0.464$ is the parameter, similar to anomalous dimension.

The new parameter $q_0$ is expressed though $M_0$ as follows:
\begin{eqnarray*}
 q_0 = \left ( \frac{M_0}{2m_c} \right )^{\frac{\nu}{1-\nu}}, \label{a4}.
\end{eqnarray*}
where 
\begin{eqnarray*}
 \nu = \mu(M_0)^{-\frac{1}{d}} \quad {\rm and } \quad 
 \mu(M_0) = \frac{1-\alpha_c}{3-\alpha_c}.
\end{eqnarray*}

\newpage

\noindent {\bf APPENDIX~2. }

For completeness sake we present here the explicit form for two--particle
distribution functions $f^{h}_{Vi}(x_V, x_1)$ in $\pi^{\pm}$ and $K^{\pm}$
mesons. 

Note, that two-particle parton distributions (like the single--particle
functions) can not be theoretically evaluated. Therefore, we use the simplest
phenomenological expression that takes into account the momentum conservation,
the $(1-x_1)^n$ dependence of sea partons, as well as the normalization
conditions as follows (see~\cite{old} for the details):
\begin{eqnarray*}
 \int_0^{(1-x_1)} f^{h}_{Vi}(x_V, x_1) d x_V = f^h_i(x_1), 
\end{eqnarray*}
where $f^h_i(x_1)$ is the single-particle distribution of an $i$--type parton
in the hadron~$h$.

For two valence quarks such a distribution has the form~\cite{old}:
\begin{eqnarray*}
 f_{VV}(x_1, x_2) = 
\frac{\Gamma(2 + \gamma_0 - \alpha_1 - \alpha_2)}
{\Gamma(1 - \alpha_1) \Gamma(1 - \alpha_2) \Gamma(\gamma_0)}
 x_1^{-\alpha_1}  x_2^{-\alpha_2} (1 - x_1 - x_2)^{\gamma_0 - 1}.
\end{eqnarray*}
For the case of one valence and one sea parton the corresponding distribution
equals:
\begin{eqnarray*}
 f_{Vj}(x_v, x_j) = N_j
\frac{\Gamma(2 + n_v - \alpha_v)} 
{\Gamma(1 - \alpha_v) \Gamma(1 + n_v)}
 x_v^{-\alpha_v}  x_j^{-1} (1 - x_v - x_j)^{n_v} (1-x_j)^k,
\end{eqnarray*}
where $N_j$ is the corresponding normalization factor of a sea parton,
$n_v = \gamma_0 - \alpha_1 - \alpha_2 + \alpha_v$, and 
$k = n_j - 1 - \gamma_0 + \alpha_1 + \alpha_2$.

The parameters of the function $f_{Vj}(x_1, x_2)$ for $\pi$ and $K$ mesons are
presented in the Table~2.

\newpage

\vspace{0.8cm}
\noindent 
\underline{ {\bf Table~1. }} 
The ratios of the total yields (at $x > 0$) of the charmed mesons in the
considered  model, the experimental results~\cite{exp3} and the Lund--model
predictions (taken also from~\cite{exp3}). 

\begin{center}
\begin{tabular}{|c|c|c|c|c|} \hline
  & $\frac{D^+ + D^-}{D^0 + \bar D^0}$ 
  & $\frac{D_s^+ + D_s^-}{D^0 + \bar D^0 + D^+ + D^-}$ 
  & $\frac{D^-}{D^+}$ & $\frac{D^0}{\bar D^0}$ \\ \hline
 our model & $0.332$  & $0.102$  & $1.16$ & $1.0$ \\ \hline 
experiment & $0.416 \pm 0.016$ & $0.129 \pm 0.012$ & $1.35 \pm 0.05$ & 
       $0.93 \pm 0.03$ \\ \hline 
Lund & $0.472$  & $0.077$ & $2.25$ & $1.09$ \\ \hline 
\end{tabular}
\end{center}


\vspace{0.8cm}
\noindent 
\underline{ {\bf Table~2. }}
Values of the parameters appearing in the functions 
$f_{Vj}(x_1, x_2) = A x_1^{-\alpha_1} x_2^{-\alpha_2} (1-x_1-x_2)^n (1-x_2)^k$.

\begin{center}

\begin{tabular}{|l|c|c|c|c|c|} \hline
 $\pi^{\pm}$ meson & $A$ & $\alpha_1$ & $\alpha_2$ & $n_v$ & $k$ \\ \hline 
  partons & & & & & \\ \hline
  $u_v d_v$         & 0.477 & 0.5 & 0.5 & 0.5 & 0 \\ \hline
  $u_v g  $         & 1.50  & 0.5 & 1.0 & 1.0 & 1.5 \\ \hline
  $u_v (u,d)_{sea}$ & 0.090 & 0.5 & 1.0 & 1.0 & 3.5 \\ \hline
  $u_v s_{sea}$     & 0.045 & 0.5 & 1.0 & 1.0 & 3.5 \\ \hline \hline
 $K^{\pm}$ meson & $A$ & $\alpha_1$ & $\alpha_2$ & $n_v$ & $k$ \\ \hline 
  partons & & & & & \\ \hline
  $u_v s_v$         & 1.27 & 0.5 & 0.0 & 0.5 & 0   \\ \hline
  $u_v g  $         & 1.34 & 0.5 & 1.0 & 1.0 & 1.5 \\ \hline
  $s_v g  $         & 3.22 & 0.0 & 1.0 & 1.5 & 1.5 \\ \hline
  $u_v (u,d)_{sea}$ & 0.08 & 0.5 & 1.0 & 1.0 & 3.5 \\ \hline
  $u_v s_{sea}$     & 0.04 & 0.5 & 1.0 & 1.0 & 3.5 \\ \hline 
  $s_v (u,d)_{sea}$ & 0.192 & 0.0 & 1.0 & 1.5 & 3.5 \\ \hline
  $s_v s_{sea}$     & 0.096 & 0.0 & 1.0 & 1.5 & 3.5 \\ \hline 
\end{tabular}
\end{center}

\newpage

\begin{figure}[t]
\begin{center}
\epsfig{file=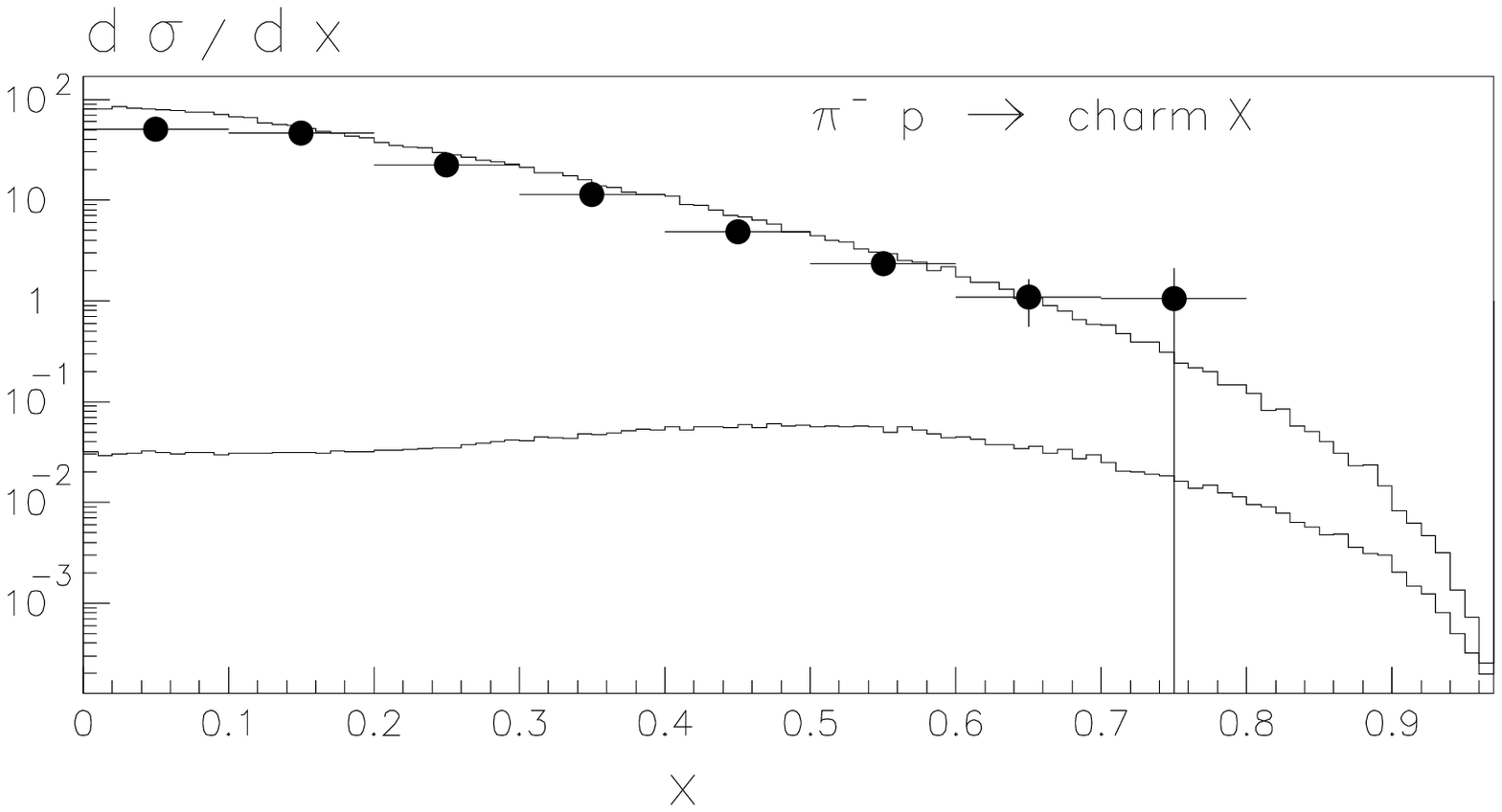,width=12cm,clip=}

\ccaption{}{
Differential distributions $\frac{d \sigma}{dx}$ for
charmed $c$ quarks for all values of $M_{c \bar c}$ (the upper histogram)
and for
$M_{c \bar c} \geq M_0 = 10$~GeV (the lower histogram). The experimental points
correspond to charmed particles yield, summed over all types of $D$ 
and $\bar D$ mesons (the reaction of $\pi^- N$ collisions 
at $E_{\pi} = 250$~GeV~\cite{exp1}.) 

}

\end{center}
\end{figure}

\newpage

\begin{figure}[t]
\begin{center}
\epsfig{file=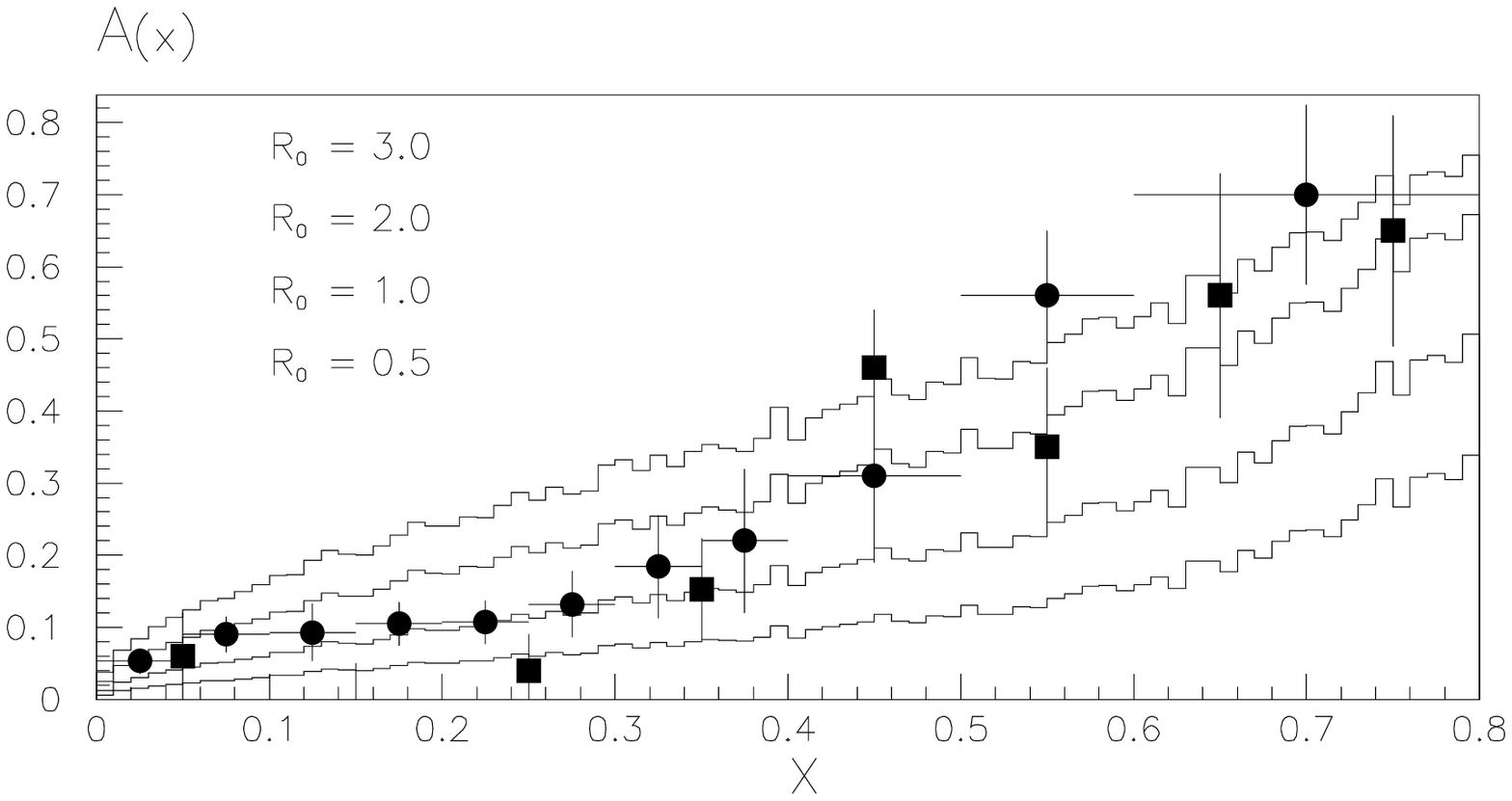,width=12cm,clip=}
\ccaption{}{
The description of the asymmetry $A(x)$ (the leading particle
effect) in $\pi^- p $ collisions~\cite{exp1,exp2} in the mechanism of the
"hard" fragmentation (see~(\ref{sig2})). The different histograms 
correspond to  different values of the parameter $R_0$ for the
recombination mechanism (see~(\ref{sig1})). The value of
$R_0 = 3.0$ corresponds to the upper histogram and so on.
}

\end{center}
\end{figure}

\newpage

\begin{figure}[t]
\begin{center}
\epsfig{file=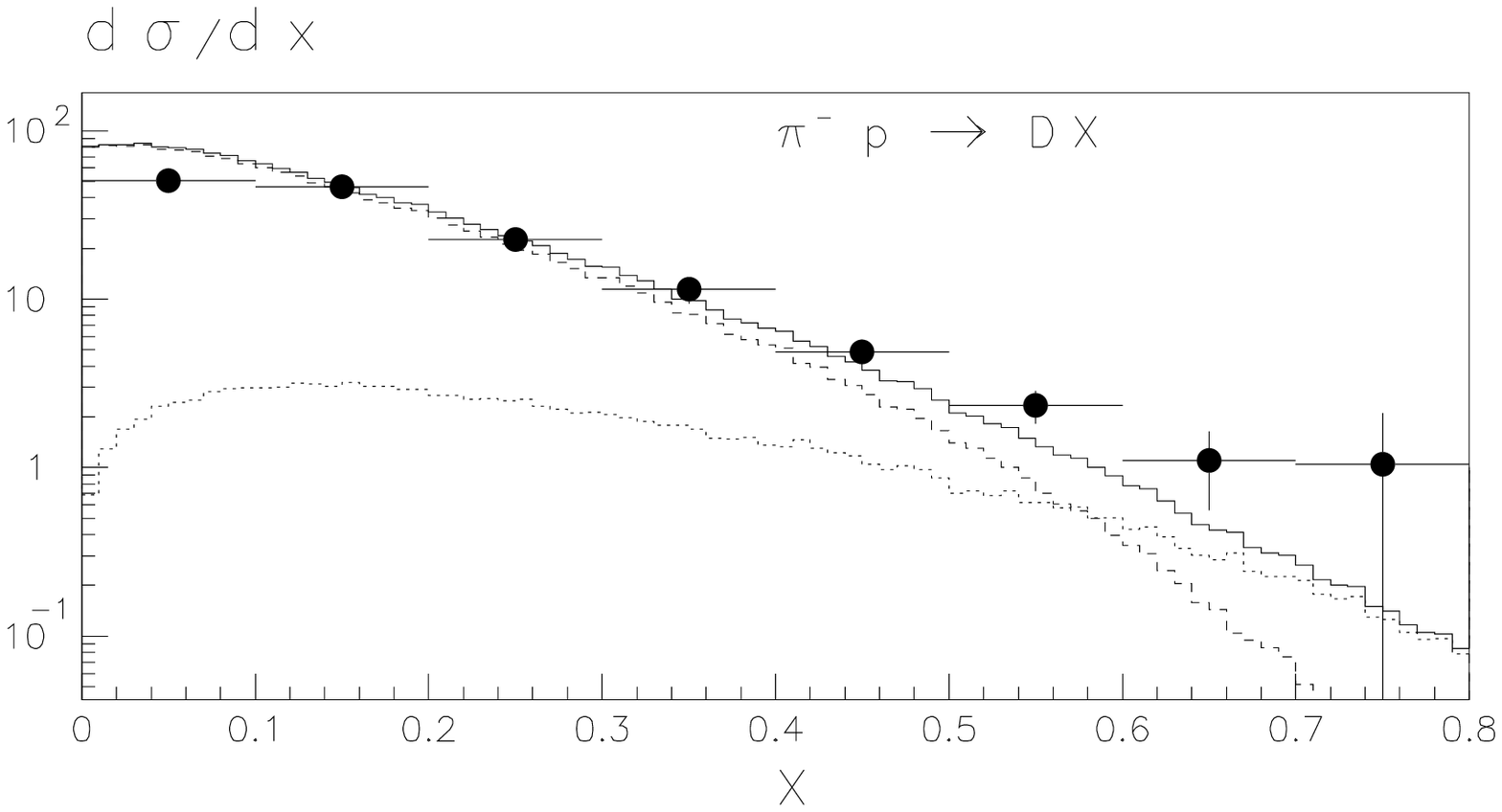,width=12cm,clip=}
\ccaption{}{
Differential distributions $\frac{d \sigma}{dx}$ for
the energy of $E_{\pi} = 250$~GeV. The experimental data are taken 
from~\cite{exp1}. 
The dotted (dashed) histogram corresponds to the recombination (fragmentation)
contribution. The solid histogram represents their sum. The cross sections are
presented in $\mu$b.

}

\end{center}
\end{figure}

\newpage

\begin{figure}[t]
\begin{center}
\epsfig{file=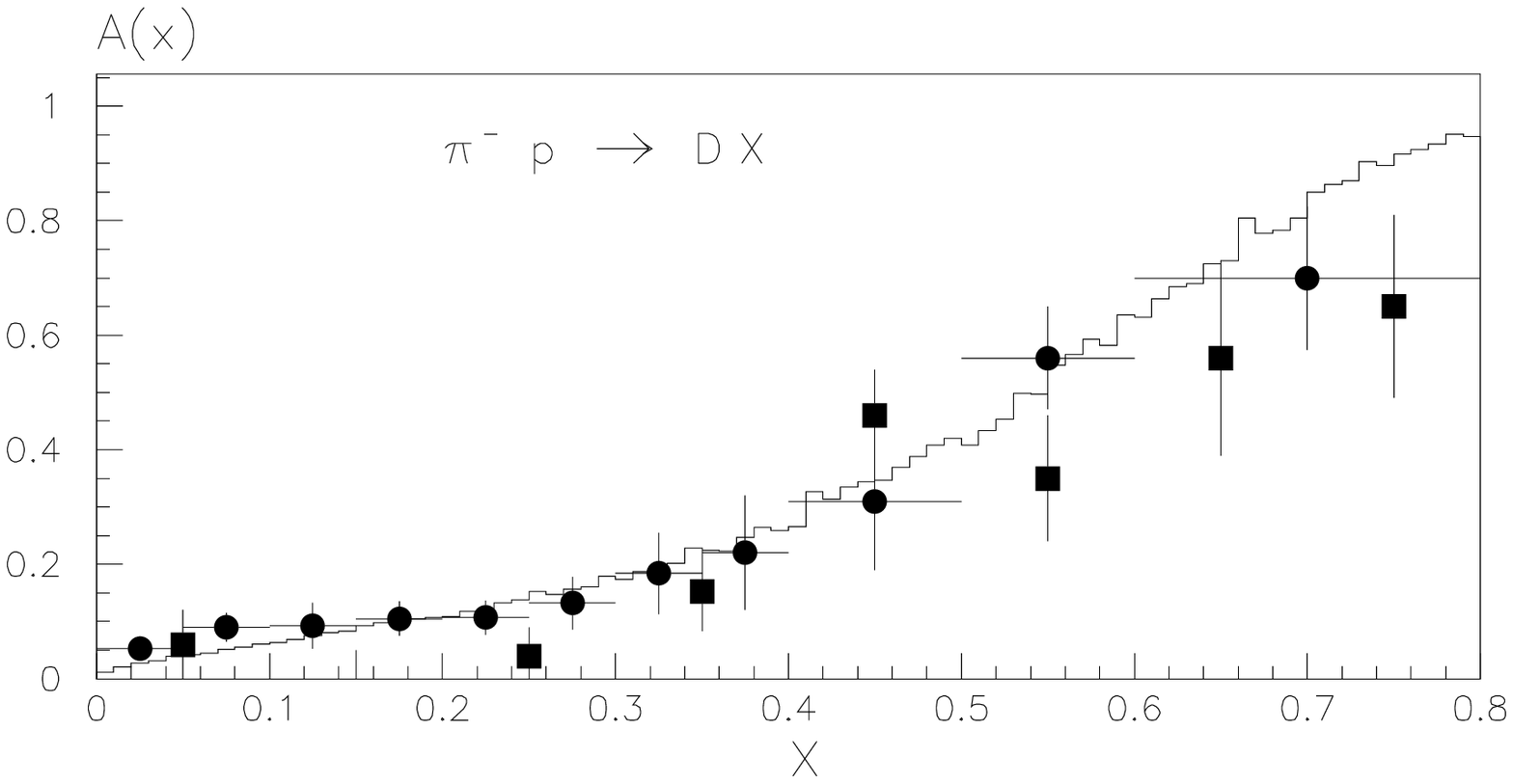,width=12cm,clip=}
\ccaption{}{
The description of the asymmetry $A(x)$ 
in $\pi^- p $ collisions~\cite{exp1,exp2} in the mechanism of the
modified fragmentation (see~(\ref{sig4})) with the help of fragmentation
function from~(\ref{a2})).

}

\end{center}
\end{figure}

\newpage

\begin{figure}[t]
\begin{center}
\epsfig{file=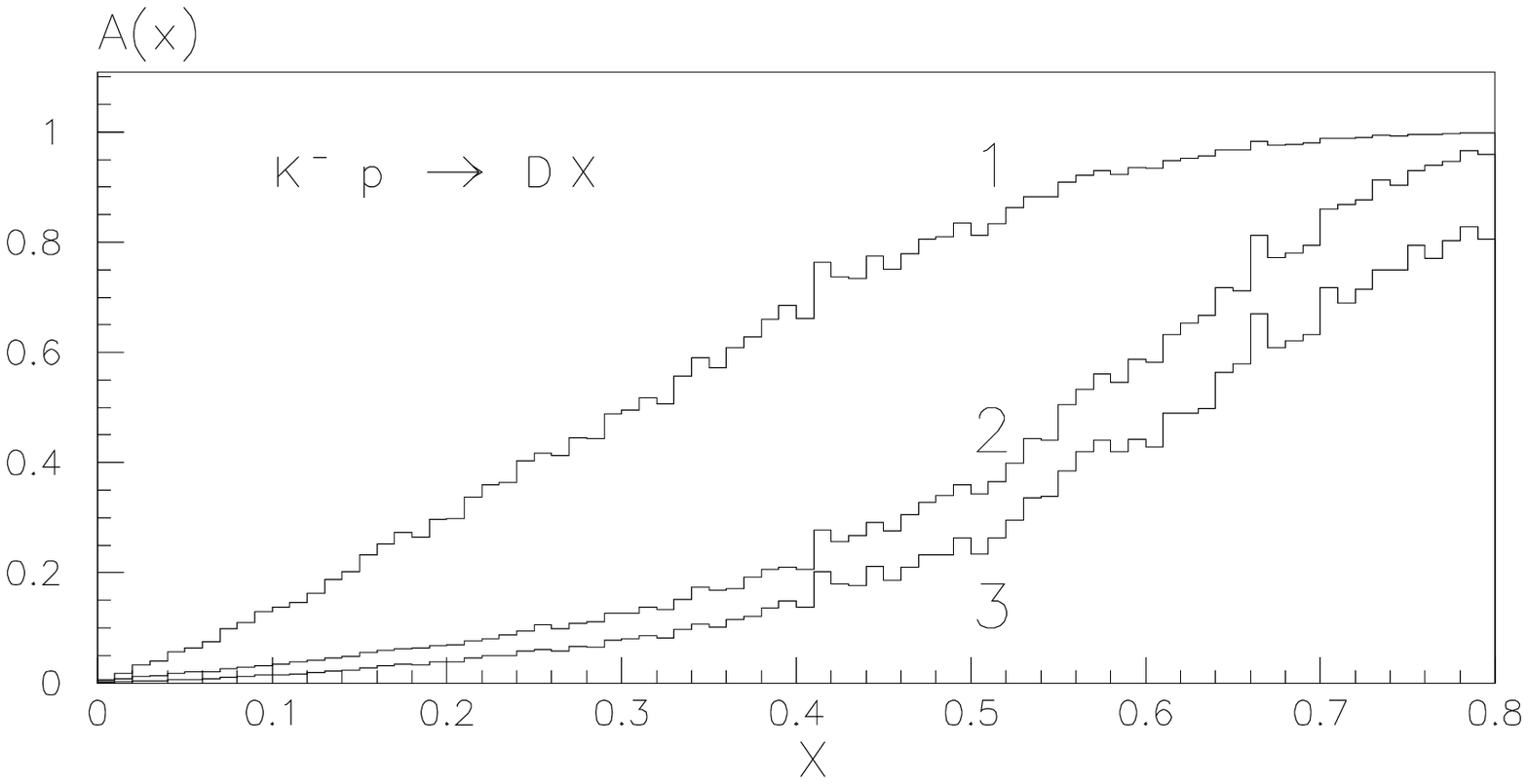,width=12cm,clip=}
\ccaption{}{
The predictions of the modified fragmentation mechanism for
the asymmetry $A(x)$ in $K^- p $ collisions at $E_K = 250$~GeV.
The histograms correspond to the ratios of: 
1)~$D_s(\bar c s)$ to $D_s(c \bar s)$ mesons, 2)~all charmed
$D + D_s$ mesons, 3)~$D_s(\bar c s)$ to all charmed $D + D_s$ mesons.

}

\end{center}
\end{figure}


\begin{thebibliography}{**}
\bibitem {1} Mangano~M., Nason~P., and Ridolfi~G., //
{\it Nucl. Phys.} {\bf B405} (1993) 507.
\bibitem {ls97} Likhoded~A.K. and Slabospitsky~S.R., //
{\it Yad. Fiz.} {\bf 60 } (1997) 1097.
\bibitem {old} Kartvelishvili~V.G., Likhoded~A.K., and Slabospitsky~S.R., // 
{\it Yad. Fiz.} {\bf 32} (1980) 236; \\ 
Kartvelishvili~V.G., Likhoded~A.K., and Slabospitsky~S.R., //
{\it Yad. Fiz.} {\bf 33} (1981) 832; \\ 
Likhoded~A.K., Slabospitsky~S.R., and Suslov~M.V. //
{\it Yad. Fiz.} {\bf 38} (1983) 727. 
\bibitem {collin} Collins, P.D.B., {\it "An Introduction to Regge Theory 
and High Energy Physics"}, Cambridge: Cambridge University Press, 1977.
\bibitem {klp} Kartvelishvili~V.G., Likhoded~A.K., and Petrov~V.A., //
{\it Phys. Lett.} {\bf B78} (1978) 615. 
\bibitem {peters} Peterson~C., Schlatter~D., Schmitt~I., and Zerwas, P., //
 {\it Phys. Rev.} {\bf D27} (1983) 105.
\bibitem {pdg} Montanet, L. {\it et al.} (Review of Particle Properties), 
{\it Phys. Rev.} {\bf D50,} Part II (1994) 1. 
\bibitem {expfr1} {\it CLEO Collaboration,} Bortoletto,~D. {\it et al.}, //
{\it Phys. Rev.} {\bf D37} (1988) 1719;
\bibitem {expfr2} {\it OPAL Collaboration,} Akers,~R. {\it et al.}, //
{\it Z.Phys.} {\bf C67} (1995) 27;  \\
{\it OPAL Collaboration,} Ackerstaff,~K. {\it et al.}, //
{\it CERN-PPE/97-093}, 1997, {\it hep--ex}/9708021; \\
{\it ALEPH Collaboration,} Buskulic,~D. {\it et al.}, //
{\it Z.Phys.} {\bf C62} (1994) 1. 
\bibitem {exp1} {\it E769 Collaboration,} Alves,~G.A. {\it et al.}, //
{\it Phys. Rev. Lett.} {\bf 77} (1996) 2392. 
\bibitem {exp2} {\it WA82 Collaboration,} Adamovich,~M. {\it et al.}, //
{\it Phys. Lett.} {\bf B305} (1993) 402; 
\newpage
{\it E769 Collaboration,} Alves,~G.A. {\it et al.}, //
{\it Phys. Rev. Lett.} {\bf 72} (1994) 812; \\ 
{\it E791 Collaboration,} Aitala,~E.M. {\it et al.}, //
{\it Phys. Lett.} {\bf B371} (1996) 157. 
\bibitem {lund} Bengtsson~H.--U. and Sj{\"{o}}strand~T. //
{\it Comput.~Phys.~Commun.} {\bf 46} (1987) 43; \\ 
Sj{\"{o}}strand~T. and Bengtsson~H.--U. // 
{\it Comput.~Phys.~Commun.} {\bf 43} (1987) 367. 
\bibitem {exp3} {\it Beatrice Collaboration,} Adamovich,~M. {\it et al.}, //
{\it Nucl. Phys.} {\bf B495} (1997) 3. 
\bibitem {batun} Batunin,~A.V., Likhoded,~A.K., and Kiselev,~V.V., //
{\it Yad. Fiz. } {\bf 49} (1989) 554.
\bibitem {expk} {\it E769 Collaboration,} Alves,~G.A. {\it et al.}, //
{\it Phys. Rev. Lett.} {\bf 77} (1996) 2388. 
\end{thebibliography}
\end{document}